\documentclass{easychair}
\usepackage{pgf}
\usepackage{xspace}

\newcommand\TCT{\ensuremath{\mathsf%
 {T\kern-0.15em\raisebox{-0.3em}{$\mathsf{C}$}\kern-0.15emT}}\xspace%
}
\newcommand{\MuTerm}{\ensuremath{\mu\textsf{Term}}\xspace}
\newcommand{\Matchbox}{\textsf{Matchbox}\xspace}
\newcommand{\TTTT}{\ensuremath{%
\mathsf{T\kern-0.2em\raisebox{-0.3em}{$\mathsf{T}$}%
\kern-0.2emT\kern-0.2em\raisebox{-0.3em}{$\mathsf{2}$}}}\xspace}
\newcommand\CETA{\textsf{C\kern-0.1exe\kern-0.4exT\kern-0.4exA}\xspace}

\newcommand{\x}[1]{\mathcal{#1}}

\input{snippets.inc}
\renewcommand{\Conid}[1]{\mathsf{#1}}
\usepackage{amsthm}
\theoremstyle{remark}
\newtheorem{remark}{Remark}

\begin{document}
\title{A Haskell Library for Term Rewriting\thanks{%
  Supported by the Austrian Science Fund (FWF): I603, J3202, P22467.}}
\titlerunning{A Haskell Library for Term Rewriting}
\author{
  Bertram Felgenhauer\inst{1}
\and 
  Martin Avanzini\inst{1}
\and
  Christian Sternagel\inst{2}
}
\authorrunning{B.~Felgenhauer, M.~Avanzini and C.~Sternagel}
\institute{
  University of Innsbruck, Austria\\
  \email{\{bertram.felgenhauer,martin.avanzini\}@uibk.ac.at}
\and
  Japan Advanced Institute of Science and Technology, Japan\\
  \email{c-sterna@jaist.ac.jp}
}
\maketitle
\begin{abstract}
We present a Haskell library for first-order term rewriting covering
basic operations on positions, terms, contexts, substitutions and
rewrite rules. This effort is motivated by the increasing number of
term rewriting tools that are written in Haskell.
\end{abstract}

\section{Introduction}
First-order term rewrite systems (TRSs)~\cite{BN98} are a simple, yet
Turing-complete
model of computation. Consequently, many interesting properties like
termination, complexity, unique normalization and confluence are
undecidable for TRSs in general. Nevertheless, many techniques for
establishing these properties have been developed and implemented in
various termination, complexity and confluence tools.
A number of these tools are written in Haskell, for example
\Matchbox~\cite{W04}, \MuTerm~\cite{L04} and \TCT~\cite{AM13}.
The certification tool \CETA~\cite{TS09} is developed using the Isabelle
proof assistent \cite{NPW02}, and uses code generation to produce a Haskell program.

Each of these tools has its own implementation of basic term rewriting
functionality. We would like to change this situation by providing a
common foundation for first-order term rewriting. To this end we have
started developing a Haskell library called \texttt{term-rewriting},
aiming to be useful and easy to use. Our hope is to turn this into
a community effort that benefits many people.
Further information is available at our website:

\centerline{
\url{http://cl-informatik.uibk.ac.at/software/haskell-rewriting/}
}

The focus of our effort is the manipulation of first-order term
rewrite systems on their own. There are different applications
for first-order rewriting in Haskell programs. For example, in
\cite{vNRHJH08} generic programming techniques are used to allow
Haskell programmers to specify transformations on algebraic data types
in the form of rewrite rules.

In Section~\ref{sec-reqs}, we discuss guiding principles and lessons
learned from previous work. Then we give an overview of the existing
library in Section~\ref{sec-impl}, and finally conclude in
Section~\ref{sec-conc}.

\section{Design}
\label{sec-reqs}
In this section, we take a brief look at the term rewriting libraries
used in \Matchbox and \TCT, and then formulate some design principles
for the \texttt{term-rewriting} library.

\subsection{Prior art}
The \Matchbox termination tool%
\footnote{\url{https://github.com/jwaldmann/matchbox}}
is based on the \texttt{haskell-tpdb} library%
\footnote{\url{https://github.com/jwaldmann/haskell-tpdb}},
which provides a comprehensive parser for the Termination Problem Database
(TPDB) XML format. At the time we discussed the library design, however,
we were not aware of this development.

Previous versions of \Matchbox used the \texttt{autolib-rewriting}
library, which is part of the \texttt{auto/*} software collection.
The library is
mature, and has been in use for a long time. It certainly covers the
basic functionality that we are looking for. The main problem is that
it pulls in a lot of dependencies from the other parts of the
\texttt{auto/lib}, which occasionally leads to type signatures that
are hard to understand:
\MatchboxSample
(In this case it turns out that $\Varid{is\_linear\_term}$ is actually
an assertion that the given term is linear, and produces an informative
message in the $\Conid{Reporter}$ monad if the check fails.)

The complexity tool \TCT is developed in conjunction with its own term
library \texttt{termlib}.%
\footnote{\url{http://cl-informatik.uibk.ac.at/software/tct/}}
Again, basic functionality is covered. Its main drawback is that the
term type is monomorphic, with variables and function symbols being
represented---essentially---as integers. In practice, this means that
virtually all code has to carry a signature holding additional
information on function symbols and variables.

Neither library is very attractive for re-use, due to their complexity and
seemingly ad-hoc design decisions. We tried to avoid this situation by
following a few basic principles to be explained in the next subsection.

\subsection{Principles}
Our main goal is to have a library that is easy to use, and useful.
We have established the following guidelines.
\begin{description}
\item[Minimal interface.]
This means foremost that each concept is represented by a single type,
when possible, and that we avoid cluttering the interface with many
variants of the same functionality without good reason.
\begin{remark}
One reviewer asked why we did not use type classes. There is ample
of room for discussion here. Maybe the most convincing reason is that
this would double the number of entities (a type class and a type
for each concept) for---in our perception---little gain.
\end{remark}
\item[Simple interface.]
The library interface is plain Haskell98, without relying on advanced
types or libraries. Data types should be as simple as possible.
\item[Consensus.] 
The implemented features should be generally useful. Put differently,
the library should not force non-obvious design decisions on the user.
This is best explained by an example. A noteworthy omission in the
current library is a type for TRSs. We do provide operations on list
of rules (Section~\ref{sec-rule}), but a TRS comes with an associated
signature, and we could not agree on what a signature should be.
\end{description}
Of course these guidelines are sometimes contradictory. Consider the
example of critical pairs. By simplicity, they should just be a pair
of terms. However, additional information like the rules that were
involved in the overlap is often required. To represent this information,
another type would be needed, violating minimality. We chose to
implement a more complex type instead (Section~\ref{sec-cp}).

\section{Implementation}
\label{sec-impl}

In this section, we give an overview of the \texttt{term-rewriting}
library implementation. The modules inhabit the $\Conid{Data.Rewriting}$
(abbreviated $\Conid{D.R}$) namespace.
As a rule, we provide one module per concept, implemented using separate
submodules for type, common operations and specialised functionality.
We rely on qualified imports to disambiguate between operations that can
be used on several types. For example, linearity is defined for terms,
rules and TRSs, so we provide functions $\Conid{D.R.Term}.\Varid{isLinear}$,
$\Conid{D.R.Rule}.\Varid{isLinear}$ and
$\Conid{D.R.Rules}.\Varid{isLinear}$. The key concepts are as follows:
\begin{description}
\item[Positions.]
A position is a list of natural numbers, each denoting an argument
position. In accordance with accessing lists in Haskell, the first
argument position has index $0$.
\PosDef
Positions can be compared in various ways (e.g.~above, below, parallel to).
\item[Terms.]
Terms are polymorphic over function symbols $f$ and variables $v$.
\TermDef
In addition to the displayed operations, we can check for ground
terms, linear terms, and whether a term is a variant or an
instance of another.

\begin{remark}
Interestingly, \texttt{haskell-tpdb} defines terms with swapped
arguments in the $\Conid{Term}$ type constructor:
Our motivation was the convention of writing $\x{T}(\x{F},\x{V})$
in the term rewriting literature, but there may be practical reasons
\NewMatchboxSample
Our motivation was the convention of writing $\x{T}(\x{F},\x{V})$
in the term rewriting literature, but there may be practical reasons
for swapping the order.
\end{remark}

\item[Substitutions.]
In the term rewriting theory, substitutions are partial functions,
usually with finite domain, from variables to terms. When applying a
substitution to a term, variables for which the substitution is
undefined are left untouched. In order to allow substitutions that
change the type of variables (which arise naturally from matching
terms), we distinguish between \emph{generalized} substitutions and the
standard kind. Attempting to apply a generalized substitution to a term
that contains a variable with undefined replacement fails; the
function yields no result, i.e., $\Conid{Nothing}$.
\SubstDef
Substitutions can be obtained by matching or unifying terms.
\MatchUnifyDef
\item[Rules.]
\label{sec-rule}
Rules are directed equations with terms on the left-hand and right-hand
sides. They are valid rewrite rules if the left-hand side is not a
variable and each variable that occurs in the right-hand side also
occurs in the left-hand side.
\RuleDef
The library supports checking of syntactical properties of rewrite rules,
e.g., left- and right-linearity. One can also determine whether a rule
is a variant or an instance of another one.
\item[Lists of rules.]
A list of rules (which is a TRS without an associated signature)
defines a rewrite relation on terms. The $\Conid{Reduct}$ type records
the position, used rule and substitution of a rewrite step in addition
to the resulting term. We can compute the reducts of a term using
a list of valid rules with respect to various strategies.
\RewriteDef
\item[Critical Pairs.]
\label{sec-cp}
Critical pairs are the reducts arising from a critical overlap of rules.
In the \texttt{term-rewriting} library, we annotate critical pairs by
the source of the two
rewrite steps (sometimes called a critical peak), the rules and the
position of the left rewrite step.
\CPDef
We also support computation of inner and outer (i.e., root) critical
pairs.
\item[Contexts.]
A context is a term with a single hole. Very few operations are
implemented for contexts.
\CtxtDef
\end{description}

In addition, the library provides a data type ($\Conid{Problem}$) and
parser for the WST (old TPDB) file format. There is also pretty
printing support for terms, rules, substitutions and problems.

\subsection{Example}
As an example, we show an implementation of the local confluence check
in the Knuth-Bendix criterion. The implementation is straightforward,
using an auxiliary function $\Varid{nf}$ that reduces a term to normal
form with respect to the given TRS.
\KnuthBendix

\section{Future}
\label{sec-conc}
We have described the current state of the \texttt{term-rewriting} Haskell
library. In our view it covers enough functionality to be useful, even
though most of it is trivial (notable exceptions are unification and the WST
parser). However, the library has very few users: There is a converter
from resource aware ML to term rewriting systems%
\footnote{\url{http://cl-informatik.uibk.ac.at/users/georg/cbr/tools/RaML/}}
and the beginnings of a confluence tool,%
\footnote{\url{https://github.com/haskell-rewriting/confluence-tool}}
but nothing more. Thus, there are likely to be omissions in the interface.

For the future, we hope that the library becomes adopted more widely.
The source code is hosted at github, so it is easy to submit bug reports
and feature requests. We also have a mailing list for users, and the
packages are available on Hackage.%
\footnote{\url{http://hackage.haskell.org/}}

There are plenty of missing features, even leaving the contentious field
of signatures aside. One interesting area is term graphs, where terms
are represented with explicit sharing. Term graphs allow for efficient
term rewriting and arise naturally as the result of unification.
Unification can also gainfully take sharing information into account.

\bibliographystyle{plain}
\bibliography{paper}
\end{document}